\documentclass[traditabstract]{aa}
\usepackage{epsfig}
\input psfig.sty
\begin{document}
\title{Forecasting cosmological constraints from age of high-$z$ galaxies}
\author{C. \ A.\ P. Bengaly Jr. \inst{1}\thanks{\email{carlosap@on.br}}
\and
M.\ A.\ Dantas \inst{2}\thanks{\email{aldinezdantas@uern.br}}
\and
J.\ C.\ Carvalho. \inst{1,3}\thanks{\email{carvalho@dfte.ufrn.br}}  
\and
J.\ S.\ Alcaniz\inst{1}\thanks{\email{alcaniz@on.br}}}
\institute{
Departamento de Astronomia, Observat\'orio Nacional, 20921-400, Rio de Janeiro - RJ, Brasil
\and
Departamento de F\'{\i}sica, Universidade Estadual do Rio Grande do Norte, 59625-620, Mossor\'o - RN, Brasil
\and
Departamento de F\'{\i}sica, Universidade Federal do Rio Grande do Norte, 59072-970, Natal - RN, Brasil}

\date{}

\abstract{We perform Monte Carlo simulations based on current age estimates of high-$z$ objects to forecast constraints on the equation of state (EoS) of the dark energy. In our analysis, we use two different EoS parameterizations, namely, the so-called Chevallier-Polarski-Linder (CPL) and its uncorrelated form, and calculate the improvements on the figure of merit (FoM) for both cases. Although there is a clear dependence of the FoM with the size and accuracy of the synthetic age samples, we find that the most substantial gain in FoM comes from a joint analysis involving age and baryon acoustic oscillation data.}


\maketitle

\section{Introduction}    

Over the last decade, a significant amount of evidence has been accumulated for the existence of a dark energy component that fuels current cosmic acceleration. This evidence comes mostly from distance measurements of type Ia supernovae (SN Ia) (Riess {\it et al.}, 1998; Permultter {\it et al.}, 1999), the baryon acoustic oscillation (BAO) feature in the large-scale distribution traced by the galaxy distribution (Peebles \& Yu, 1970; Blake \& Glazebrook, 2003; Eisenstein {\it{et al.}}, 2005), and measurements of the cosmic microwave background (CMB) anisotropies (Spergel {\it{et al.}}, 2005; Ade {\it{et al.}}, 2013). Together, these results provide strong support for the standard cosmological scenario and an interesting link connecting the inflationary flatness prediction with current astronomical observations.

Another important class of evidence comes from estimates of the age of the Universe. In reality, since the days of pre-dark energy, this kind of observation has been one of the most pressing pieces of data supporting the idea of a late-time cosmic acceleration  (see, e.g., Krauss \& Turner 1995; Bolte \& Hogan 1995; Dunlop {\it{et al.}} 1996;  Alcaniz \& Lima, 1999; Jimenez \& Loeb 2002). In this regard, age estimates of high-$z$ objects provide effective constraints on cosmological parameters since the evolution of the age of the Universe differs from scenario to scenario, which means that models that are able to explain the total expanding age may not be compatible with age estimates of high-$z$ objects (see Fria\c{c}a {\it{et al.}} 2005; Dantas {\it{et al.}} 2007; 2011). This kind of analysis, therefore, is particularly interesting and complementary to those mentioned ealier, which are essentially based on distance measurements to a particular class of objects or physical rulers (see Jimenez \& Loeb 2002 for discussion on a cosmological test based on relative galaxy ages).

Our goal in this {\it{Research Note}} is to investigate the constraining power of future age data on the parameters of the dark energy equation of state (EoS). To this end, we assume the observational error distribution of a sample of 32 passively evolving galaxies studied by Simon {\it et al.} (2005) and run Monte Carlo simulations to generate synthetic samples of galaxy ages with different sizes and characteristics. To perform our analysis we assume the so-called Chevalier-Polarski-Linder (CPL) (Chevalier \& Polarski, 2001; Linder, 2003) dark energy EoS parameterization and its uncorrelated form (Wang, 2008). We discuss the improvement in the figure of merit (FoM) for the EoS parameters of both parameterizations with the size and precision of age samples as well as with the combination of age data and current baryonic acoustic oscillation (BAO) measurements.

\begin{figure*}[t]
	\label{fig:fig1}
	\includegraphics[width = 4.5cm, height = 4.36cm, angle = -90]{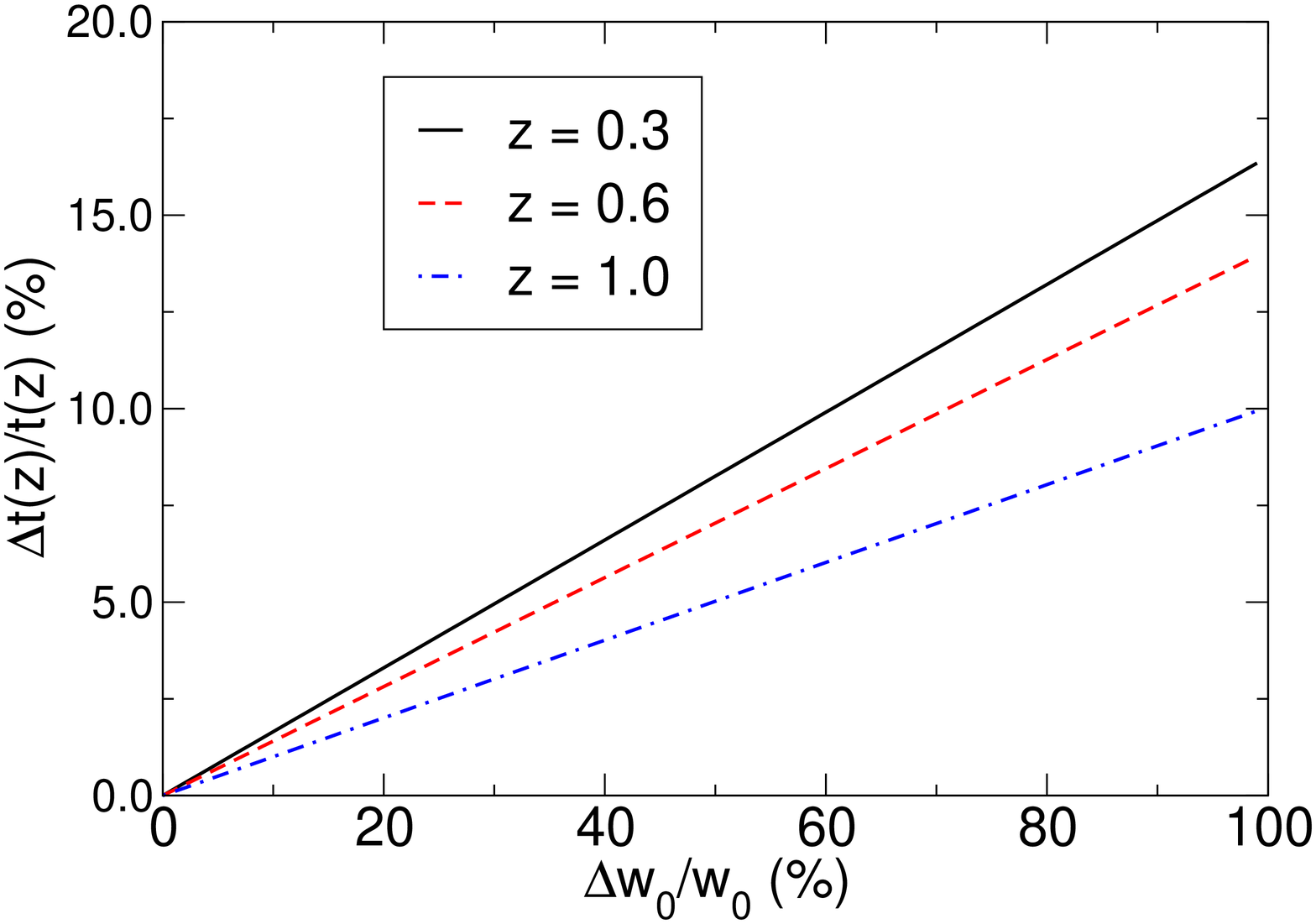}
	\hspace{0.002cm}
	\includegraphics[width = 4.5cm, height = 4.36cm, angle = -90]{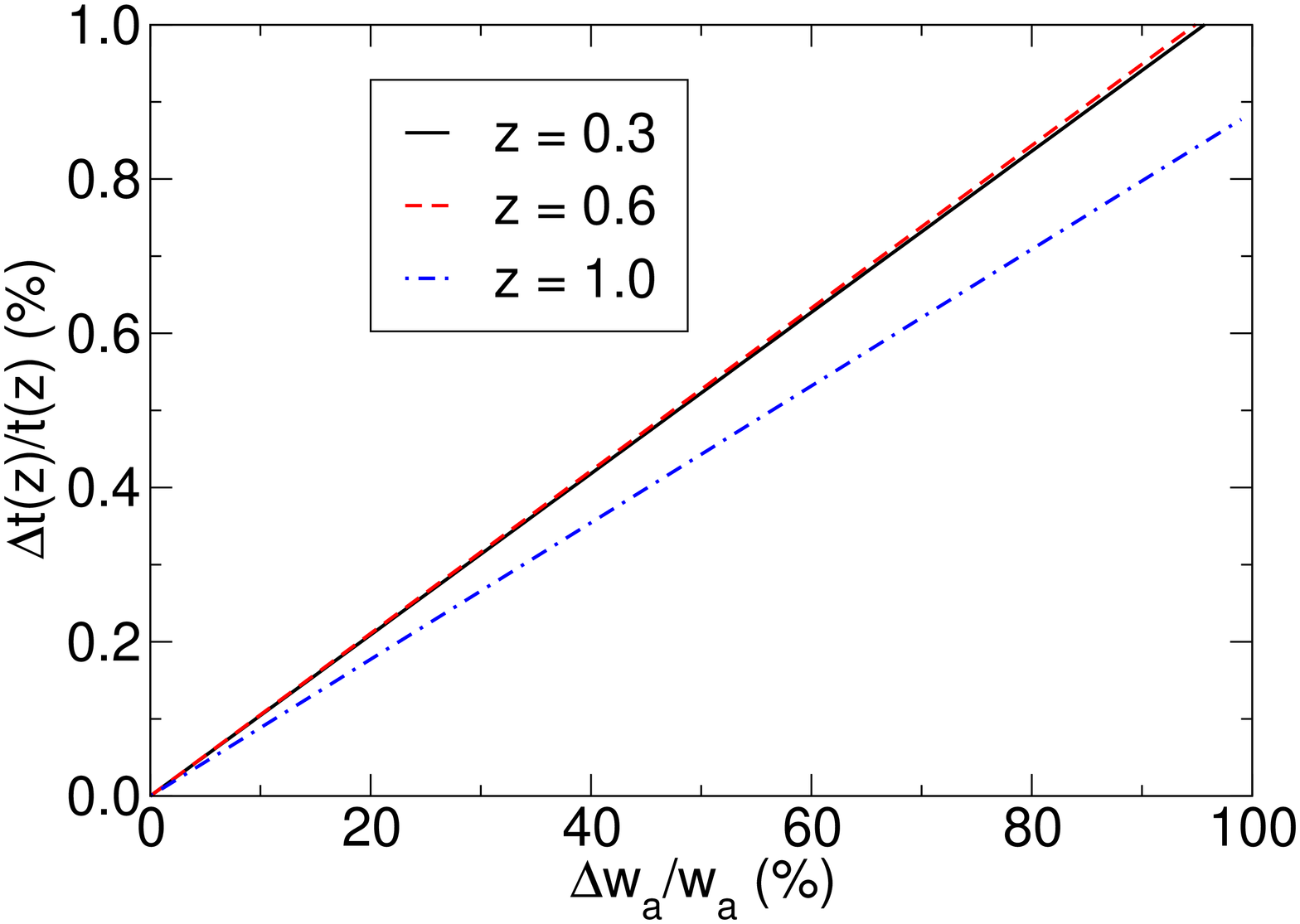}
	\hspace{0.002cm}
	\includegraphics[width = 4.5cm, height = 4.36cm, angle = -90]{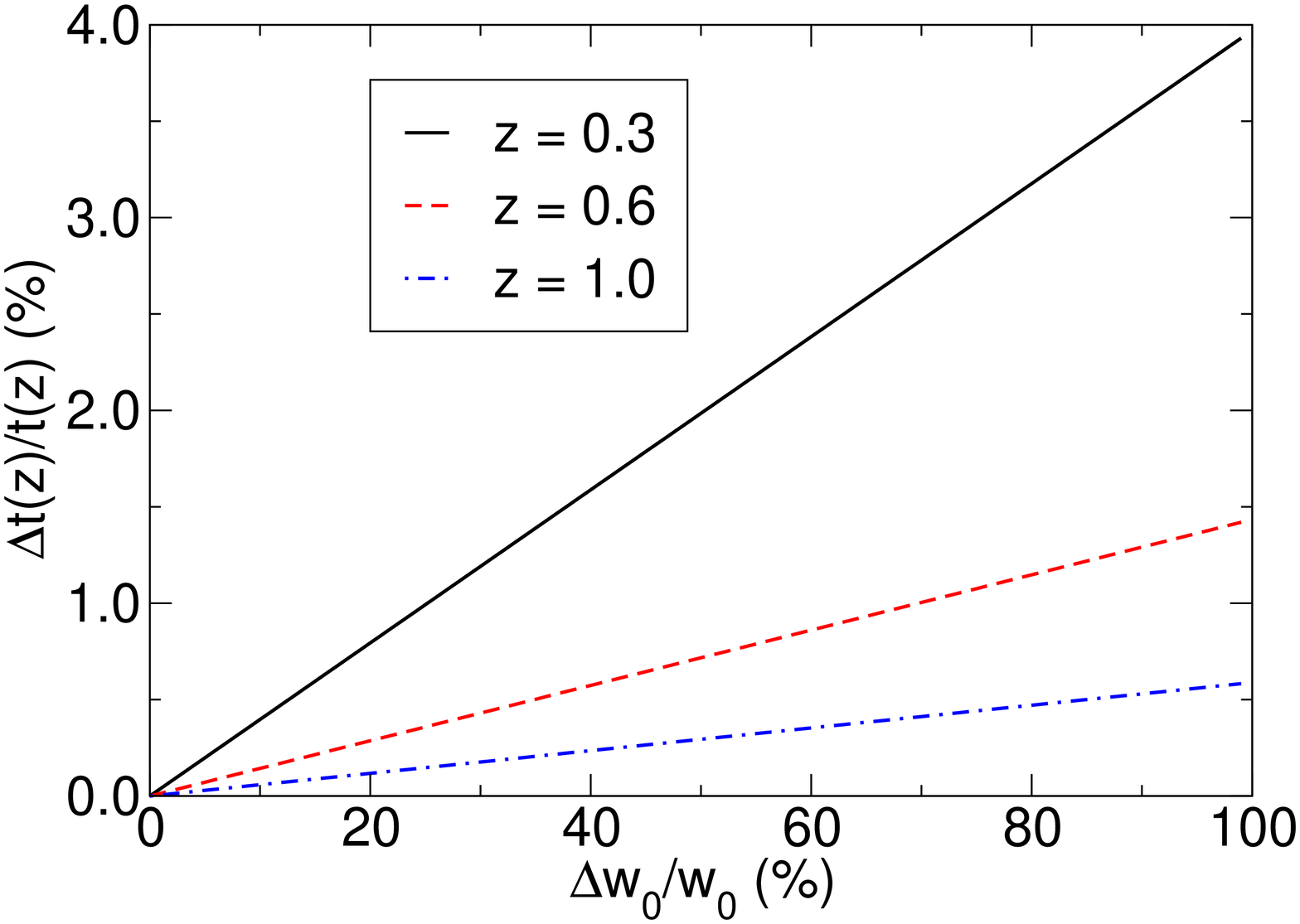}
	\hspace{0.002cm}
	\includegraphics[width = 4.5cm, height = 4.36cm, angle = -90]{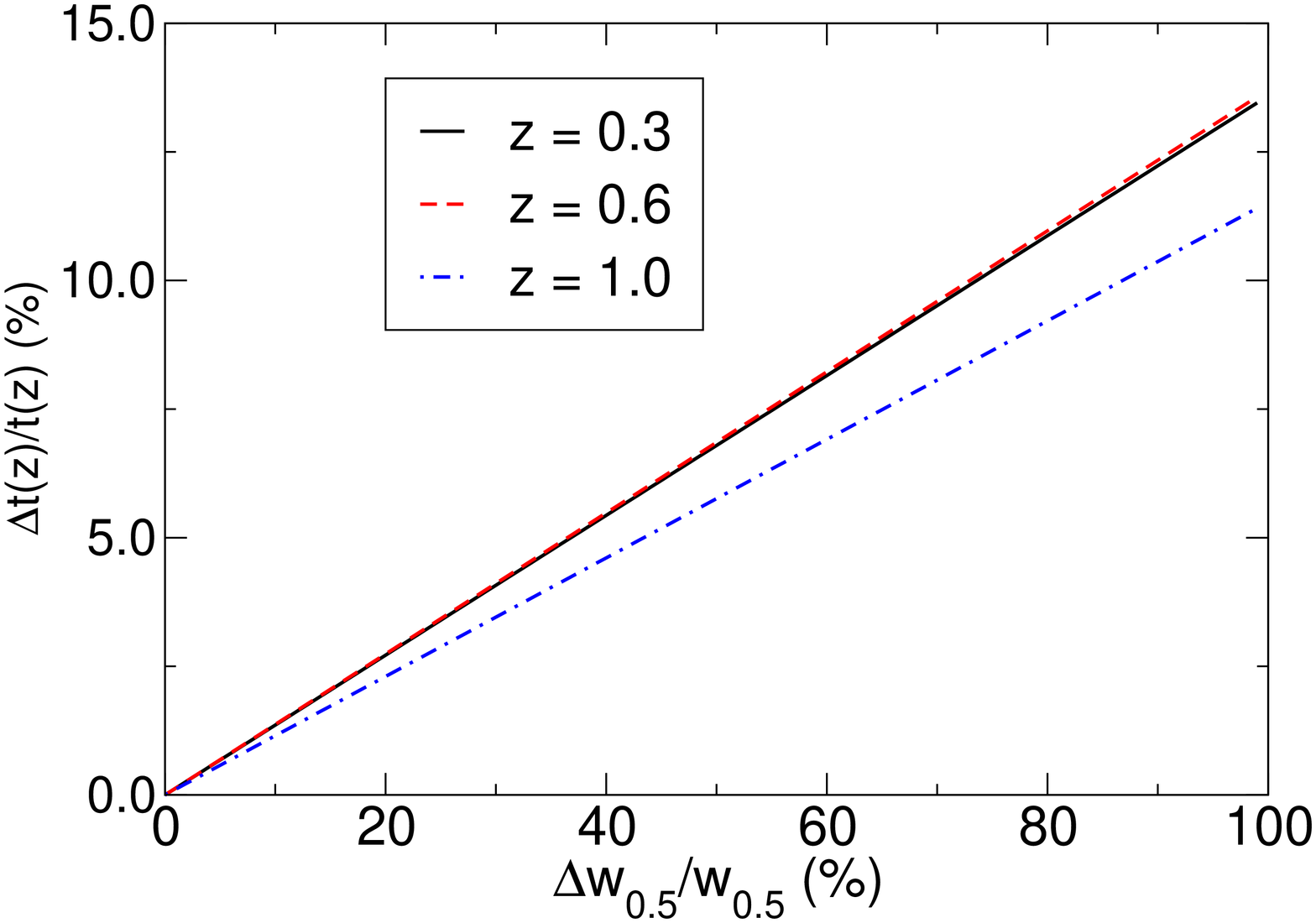}
	\caption{$\Delta t(z)/t(z)$ versus $\Delta p_i/p_i$ assuming the model parameters discussed in the text for three values of $z$.}
\end{figure*}

\begin{figure*}%[t]
	\label{fig:fig3}
	\includegraphics[width = 5.5cm, height = 6.0cm, angle = -90]{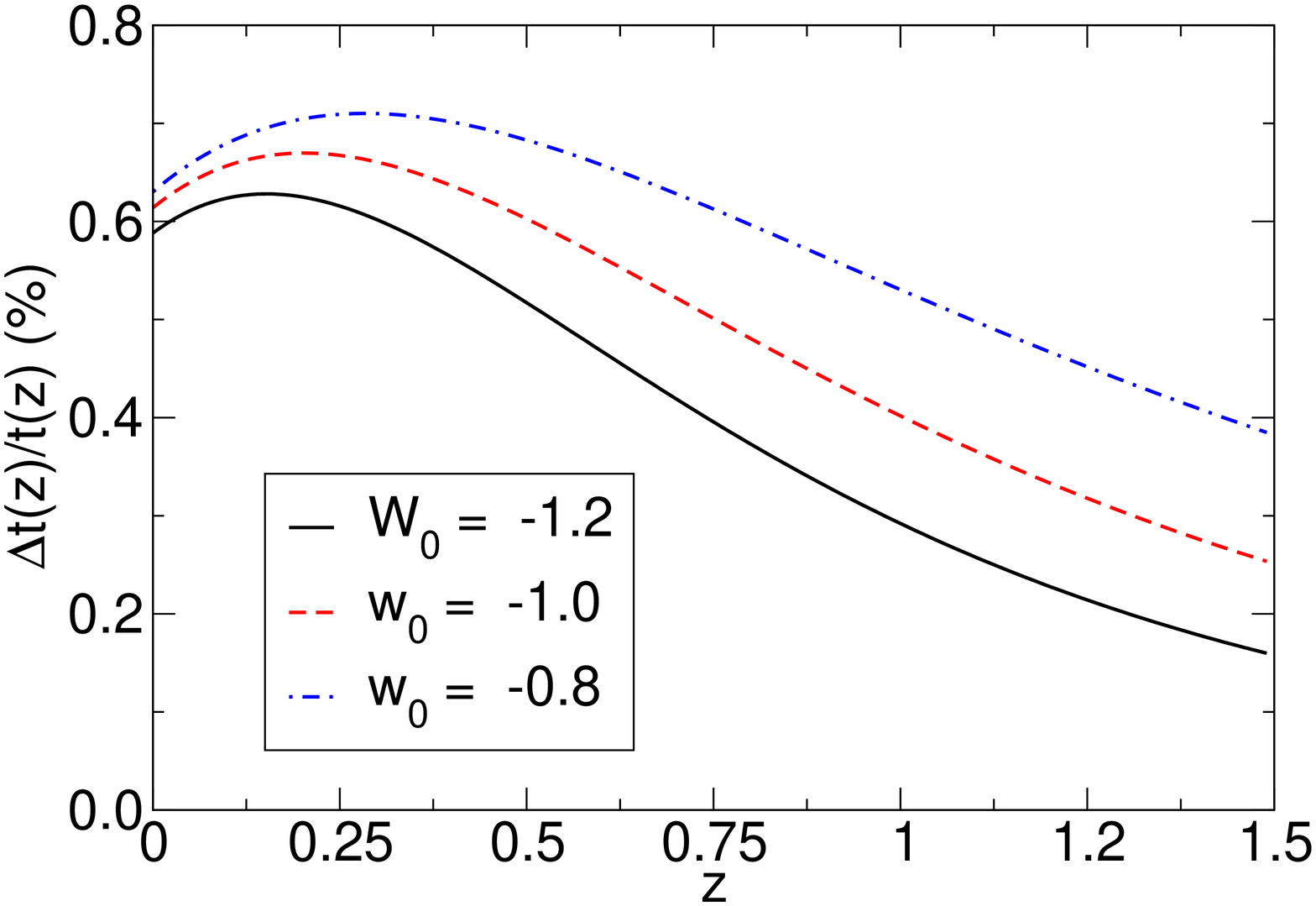}
	\hspace{0.1cm}
	\includegraphics[width = 5.5cm, height = 6.0cm, angle = -90]{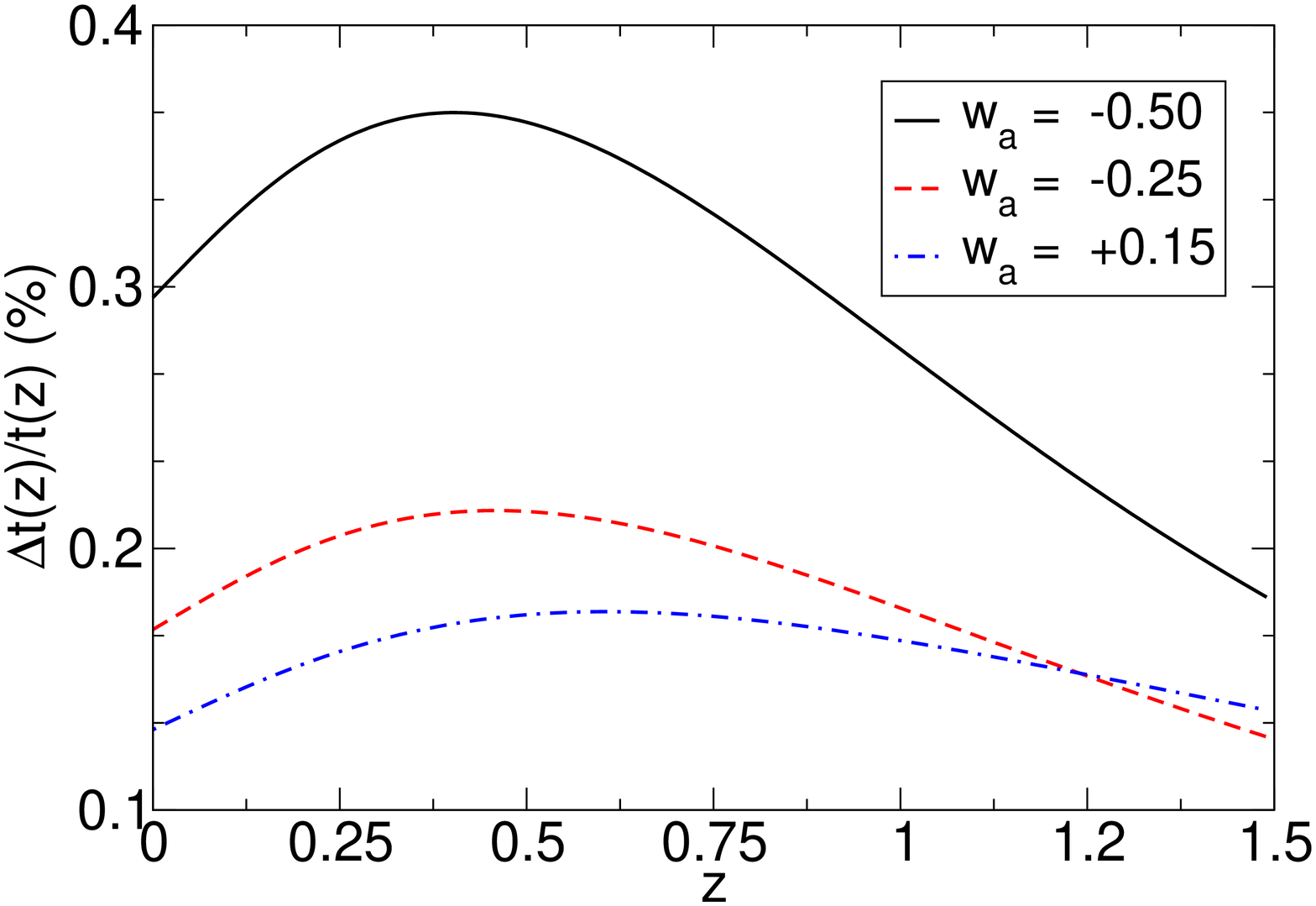}
	\hspace{0.1cm}
	\includegraphics[width = 5.5cm, height = 6.0cm, angle = -90]{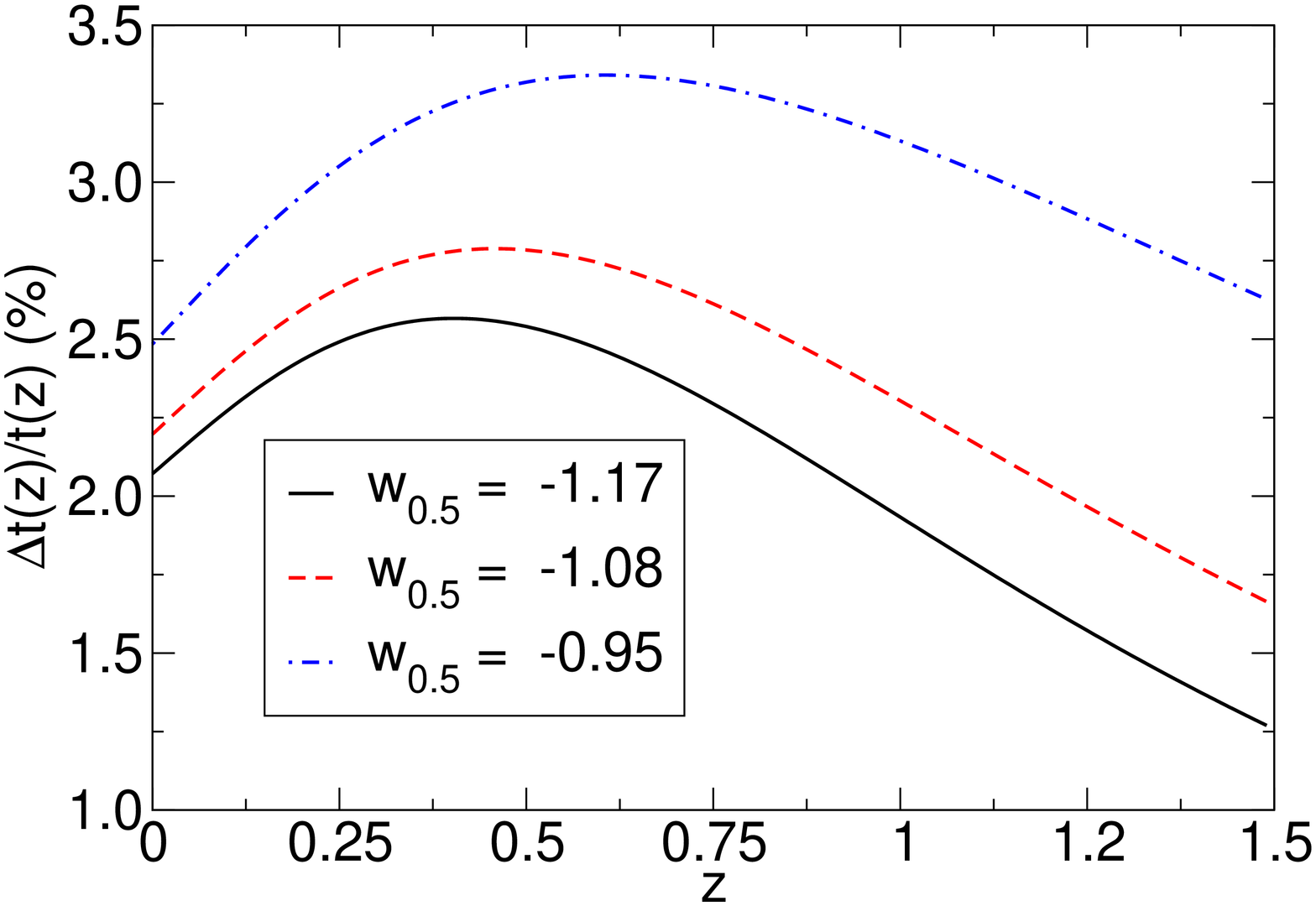}
	\caption{$\Delta t(z)/t(z)$ versus $z$ for some values of $w_0$ and $w_{\rm{a}}$ and $w_{0.5}$.}
\end{figure*}

%========================== Age-z relation =================================  

\section{The age-redshift relation}

In our analyses we consider a flat universe dominated by non-relativistic matter (baryonic and dark) and a dark energy component. In this background, the theoretical age-$z$ relation $t(z_{\rm{i}})$ of an object at redshift $z_{\rm{i}}$ can be written as (Sandage, 1988; Peebles 1993)
\begin{eqnarray} 
  \label{eq:t_z}
  t(z_{\rm{i}}, \mathbf{\theta}) = \int_{z_i}^{\infty}{\frac{dz'}{(1+z'){{h}(z',\mathbf{\theta})}}}\;,
\end{eqnarray}
where $\mathbf{\theta}$ stands for the parameters of the cosmological model under consideration and ${h}(z,\mathbf{\theta})$ is the normalized Hubble parameter, given by 
\begin{eqnarray}
	\label{eq:H(z)}
	{h(z)} = {{\sqrt{\Omega_{\rm{m}}(1+z)^3 + (1-\Omega_{\rm{m}})f(z)}}},
\end{eqnarray}
with 
\begin{equation}
\label{fz}
f(z)= 3\int_0^z \frac{1+w_{\rm{DE}}(z')}{1+z'}dz'\;. 
\end{equation}
For the dark energy EoS, $w_{\rm{DE}}$, we consider the CPL parametrization 
\begin{equation}
\label{eq:w_CPL}
w_{\rm{DE}}(a) = w_{\rm{0}} + w_{\rm{a}}(1-a)\;.
\end{equation}
Wang (2008) derived an uncorrelated form for the above parameterization by rewriting it at the value $a_c$ (or, equivalently, $z_c$) at which the parameters $w_0$ and $w_{\rm{a}}$ are uncorrelated, i.e.,
\begin{equation}
\label{paramW}
 w_{\rm{DE}}(a) = \left(\frac{a_c - a}{a_c -1}\right)w_0 + \left(\frac{a - 1}{a_c - 1}\right)w_{z_{c}}\;.
\end{equation}
In our analyses, we follow Wang (2008) and consider $a_c = 2/3$ ($z_c = 0.5$), so that the above equation can be written in terms of $z$ as
\begin{equation}
\label{eq:w_c}
w_{\rm{DE}}(z) = 3w_{0.5}-2w_0 + \frac{3(w_0 - w_{0.5})}{1+z}\;,
\end{equation}
where $w_{0.5} \equiv w(z = 0.5)$.  As mentioned earlier, Eq. (\ref{eq:w_c}) is a rearrangement of parametrization (\ref{eq:w_CPL}) that minimizes the correlation between the parameters $w_0$ and $w_{\rm{a}}$ and allows us to obtain tighter constraints on the parametric space. The parameters $w_{0.5}$, $w_0$, and $w_a$ are directly related by $w_{0.5} = w_0 + \frac{w_a}{3}$ (see Wang (2008) and Sendra \& Lazkoz (2012) for more details). 

From the above equations, we calculate the relative error in the expansion age as a function of the relative error in the EoS parameters from $\Delta t^2 = ({\frac{\partial t}{\partial {w_i}}})^2\delta {w_i}^2$, where $w_i$ stands for $w_0, w_a$ and $w_{0.5}$. Neglecting errors on $z$ and fixing $\Omega_{\rm{m}}$ = 0.27, $w_0$ = -1.0, $w_{\rm{a}}$ = -0.25 (Panels a and b) and, equivalently, $w_{0.5} = -1.08$ (Panels c and d), Fig. 1 shows $\Delta t(z)/t(z)$ versus $\Delta w_{{i}}/w_{{i}}$ for $z =0.3, 0.6$, and $1.0$. Panels (a) and (b) refer to the CPL parameterization, where we note only a slight dependence of $\Delta t(z)/t(z)$ with redshift. In order to constrain $w_0$ at a 10\% level, there must be an accuracy for $\Delta t(z)/t(z)$ of 1.65\% at $z = 0.3$ and of 1\% at $z = 1.0$. We also note that much better measurements are required to constrain $w_{\rm{a}}$ at a level of 20\%. In this case, we estimate $\Delta t(z)/t(z) \simeq 0.20\%$ for $z$ = 0.3 and 0.6, and $\Delta t(z)/t(z) \simeq 0.18\%$ for $z$ = 1.0, which 
are beyond the accuracy expected in current planned observations (see, e.g., Simon {\it{et al.}} 2005; Crawford {\it{et al.}} 2010).  

We also performed the same analysis for parametrization (\ref{eq:w_c}), displayed in Panels (c) and (d). Compared to the previous case, we note that the accuracy required to measure $w_0$ at a 10\% level should be improved by a factor of 5, whereas to obtain a 20\% measurement of $w_{0.5}$ the accuracy of $\Delta t(z)/t(z)$ could be reduced by a factor of 25. This clearly shows the effectiveness of $t(z)$ data in measuring the parameter $w_{0.5}$, which is also directly related to the time-dependent part of the dark energy EoS. For completeness, we also show the dependence of $\Delta t(z)/t(z)$  with redshift for some selected values of $w_i$ (Fig. 2). For these values, the curve $\Delta t(z)/t(z)$ presents a maximum at low-$z$, which indicates that age data at this redshift interval must impose more restrictive bounds on $w_i$ than those at high-$z$.

\begin{figure*}[] 
	\includegraphics[width = 5.5cm, height = 6.0cm, angle = -90]{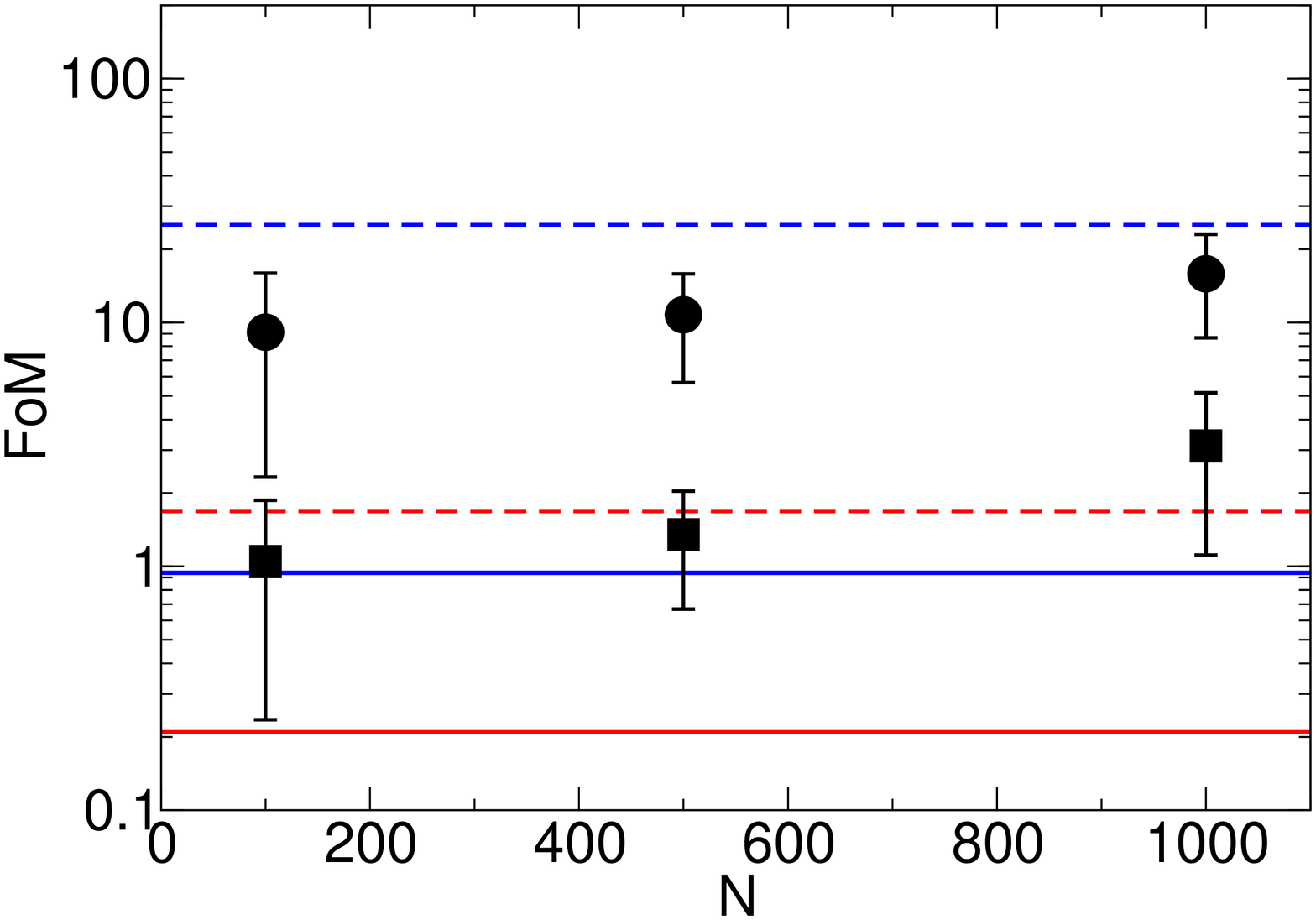}
	\hspace{0.1cm}
	\includegraphics[width = 5.5cm, height = 6.0cm, angle = -90]{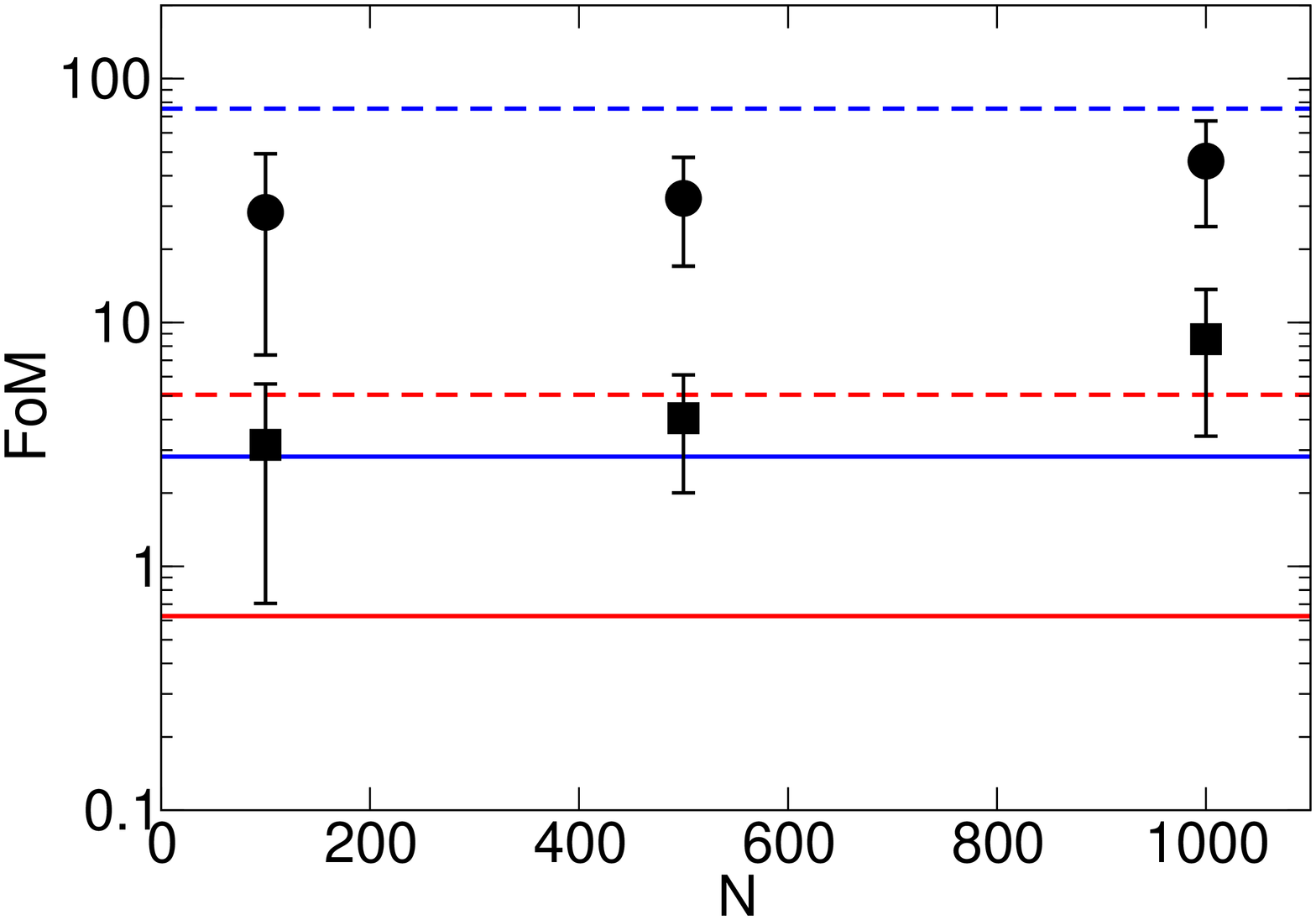}
	\hspace{0.1cm}
	\includegraphics[width = 5.5cm, height = 6.0cm, angle = -90]{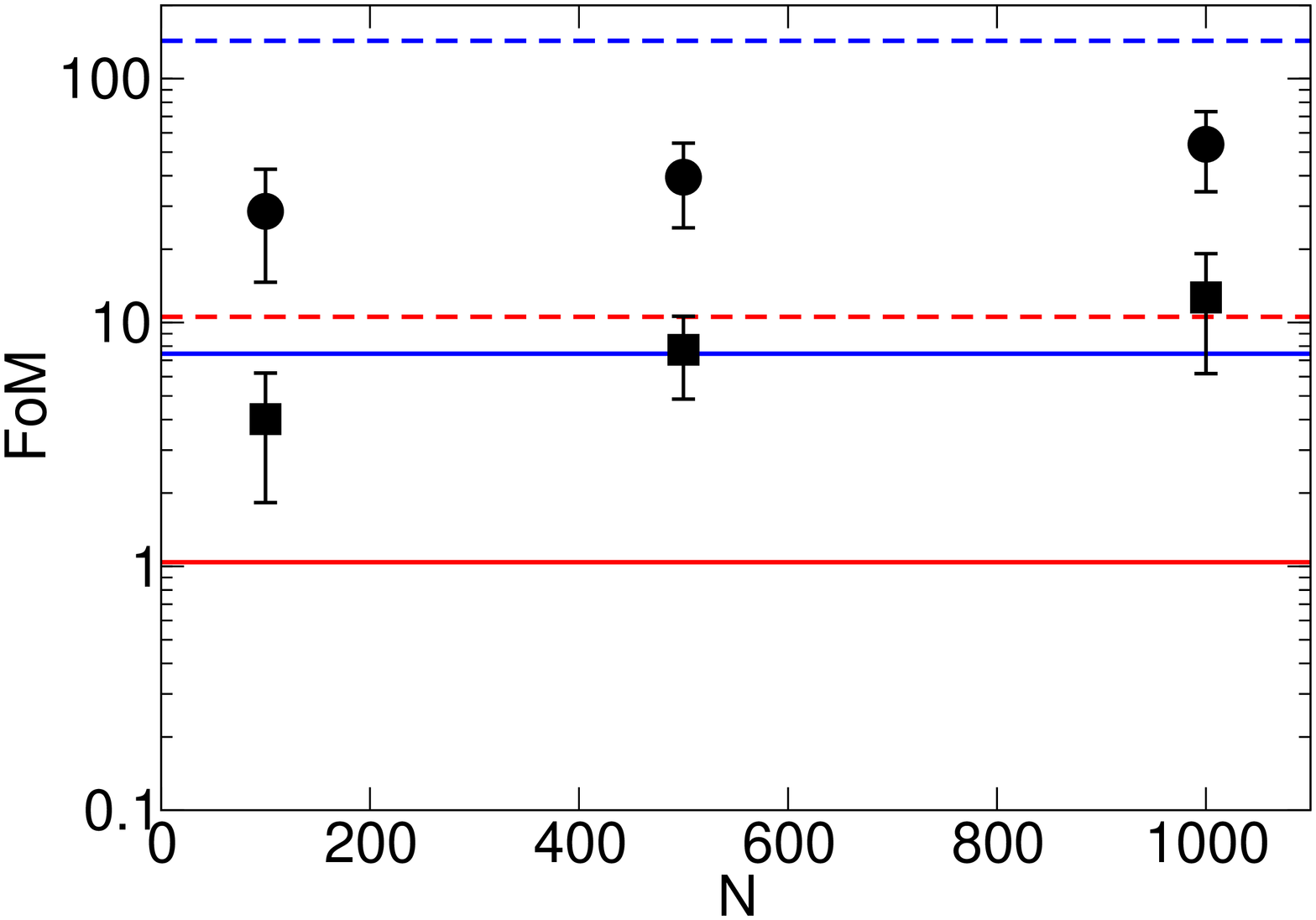}
	\caption{Figure of merit as a function of the number of data points $N$ for parametrization (\ref{eq:w_CPL}) (Panel a) and (\ref{eq:w_c}) (Panel b) assuming $\sigma_t = 10\%$. Solid squares correspond to the age simulated data only whereas solid circles stand for a joint analysis with the BAO data described in the text. Solid (dashed) red lines represent the FoM for the current observational $t(z)$ ($t(z)$ + BAO) data while solid (dashed) blue lines represent the FoM obtained from the Union2.1 compilation (SNe Ia + BAO).}
	\label{fig:fom_sigma10}
\end{figure*}

%FoM x N ($\sigma_t$ = 5\%)
\begin{figure*}[] 
	\includegraphics[width = 5.5cm, height = 6.0cm, angle = -90]{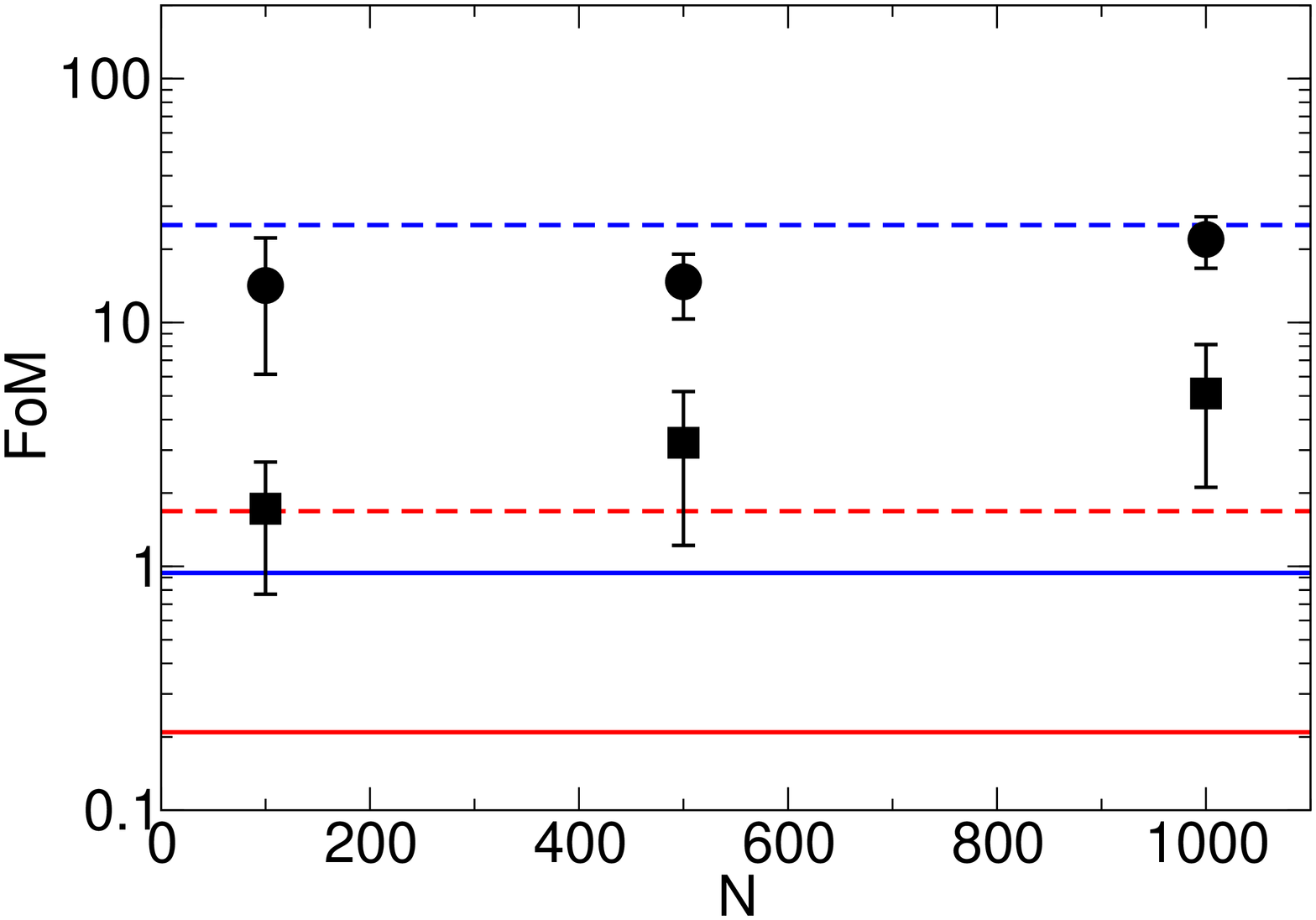}
	\hspace{0.1cm}
	\includegraphics[width = 5.5cm, height = 6.0cm, angle = -90]{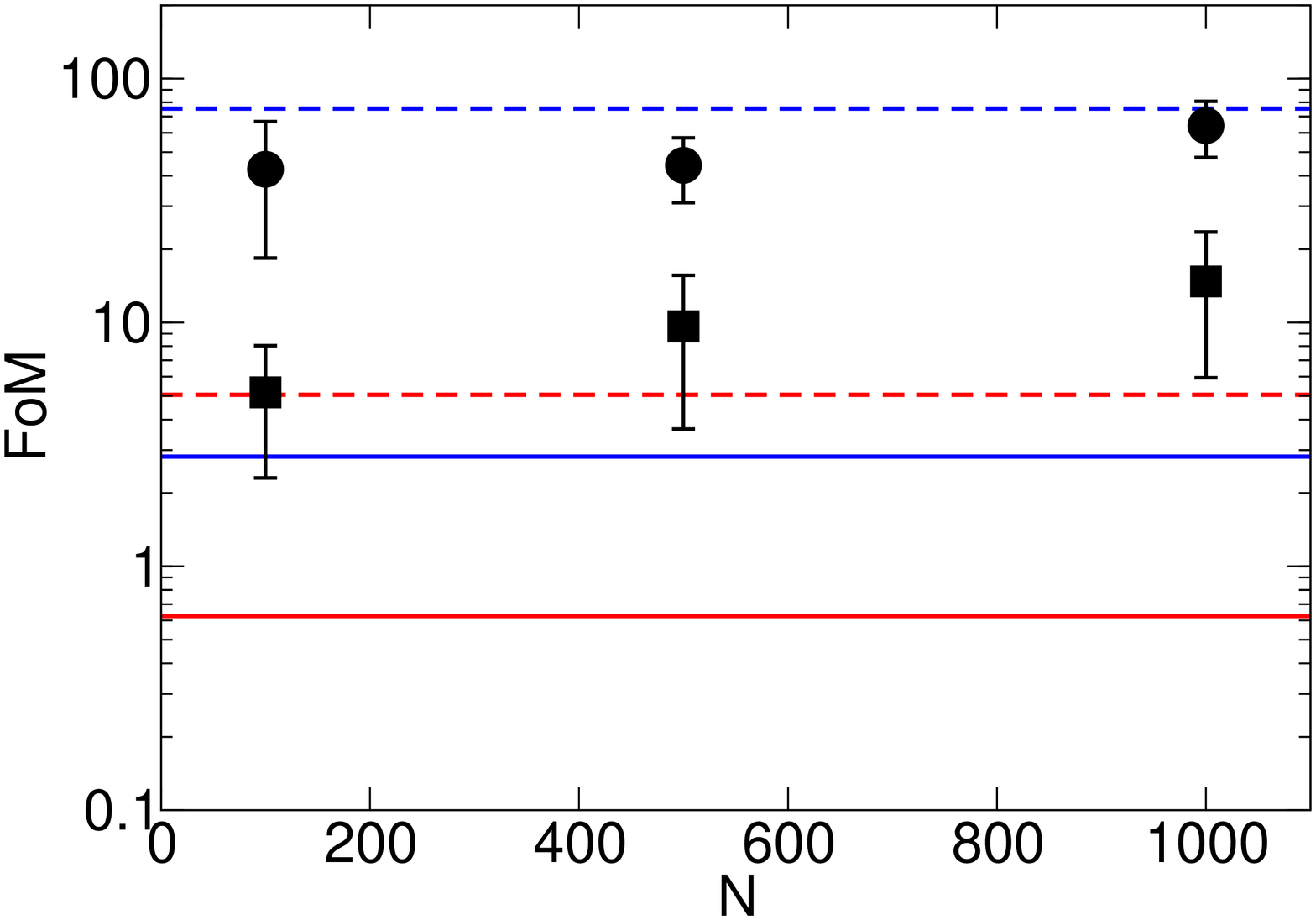}
	\hspace{0.1cm}
	\includegraphics[width = 5.5cm, height = 6.0cm, angle = -90]{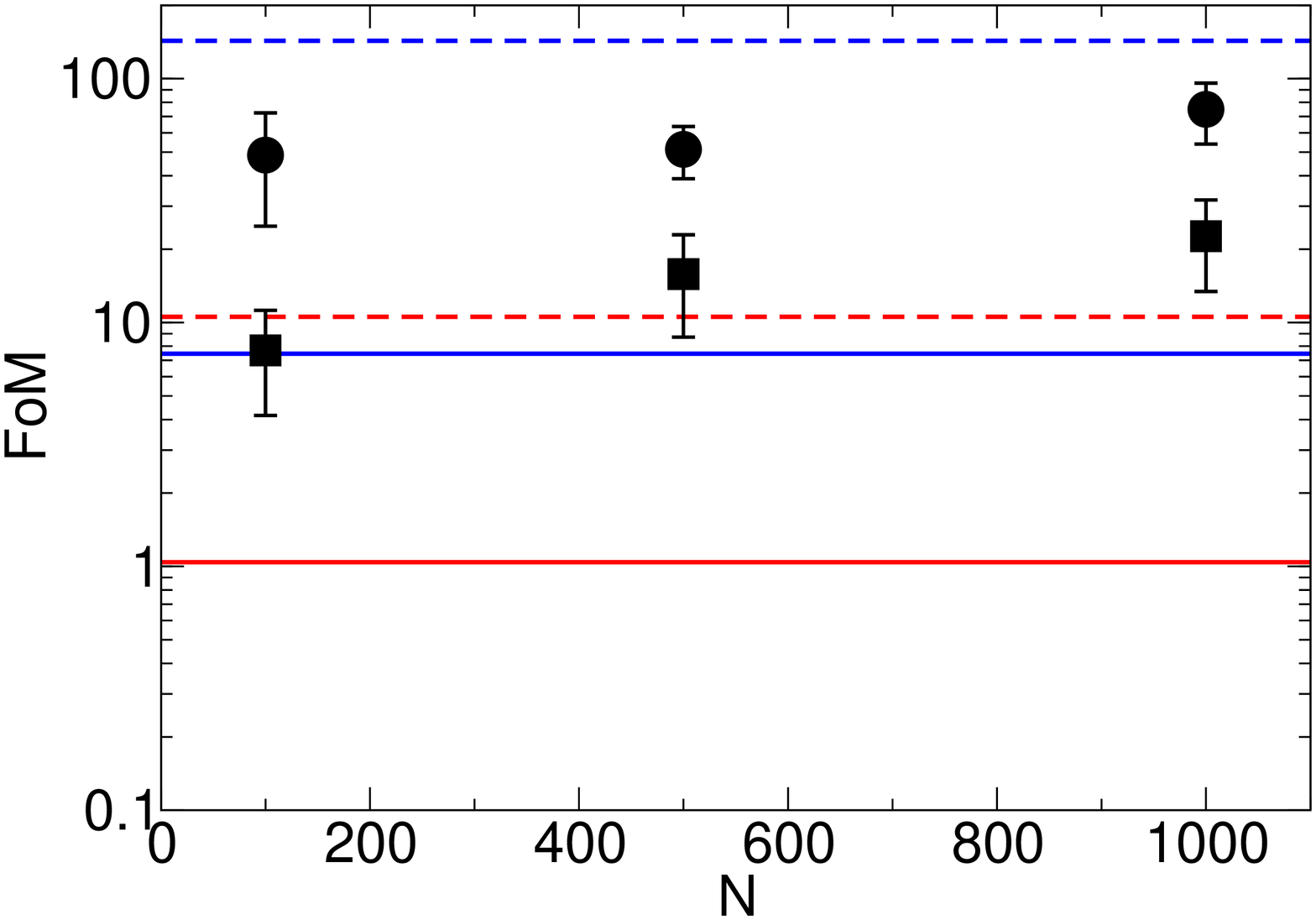}
	\caption{The same as Fig. 3 for $\sigma_t = 5\%$.}
	\label{fig:fom_sigma05}
\end{figure*}

\section{Numerical simulations}

We perform Monte Carlo (MC) simulations to generate $t(z)$ samples with different sizes and accuracy and study the expected improvement on the FoM for parameterizations (\ref{eq:w_CPL})-(\ref{eq:w_c}). Our simulations assume the current observational error distribution ($\sigma_t = 10\%$) of the $t(z)$ data given by Simon {\it{et al.}} (2005), which consist of 32 old passively evolving galaxies distributed over the redshift interval $0.11 \leq z \leq 1.84$. We then use a normal distribution centered at the $t(z_{\rm{i}})$ prediction of the chosen fiducial model, namely, a spatially flat Lambda Cold Dark Matter ($\Lambda$CDM) model with $\Omega_m = 0.27$ and $H_0$ = 74.3 $\pm$ 3.6 km/s/Mpc, which is consistent with current data from CMB (Komatsu {\it et al.}, 2010) and differential measurements of Cepheids variable observations (Riess {\it et al.}, 2009).

According to some authors (see, e.g., Simon {\it{et al.}} 2005; Crawford {\it{et al.}} 2010), future observations of passively evolving galaxies will be able to provide age estimates with $\sigma_t \leq 10\%$. In our simulations, therefore, we adopt two values of  $\sigma_t$, i.e.,  $\sigma_t = 5\%$ and $\sigma_t = 10\%$, and divide our samples into groups of 100, 500 and 1000 data points evenly spaced in the redshift range $0.1 \leq z \leq 1.5$. This makes it possible to study the expected improvement of the FoM not only as a function of the number of objects $N$, but also as a function of the precision of future cosmological observations. %For the sake of illustration, one Monte Carlo realization of 100 values of $t(z)$ with $\sigma_t = 10\%$ is shown in Fig. 3.

\begin{table}%[!t]
\begin{center}
\label{tab_union21_svj_fom_errors_rho_z} 
\begin{tabular}{llllllll}
\hline 
Parametrization (\ref{eq:w_CPL})\\  
Data & $\sigma(w_{0})$ & $\sigma(w_{\rm{a}})$ & FoM \\
\hline
Age & 1.92 & 9.88 & 0.21 \\
Age + BAO & 0.55 & 3.62 & 1.68 \\ 
Union2.1 & 1.41 & 4.63 & 0.4 \\
Union2.1 + BAO & 0.18 & 1.01 & 13.75 \\
\hline
Parametrization (\ref{eq:w_c}) \\
Data & $\sigma(w_{\rm 0})$ & $\sigma(w_{\rm 0.5})$  & FoM \\
\hline
Age & 1.92 & 1.52 & 0.62 \\ 
Age + BAO & 0.55 & 0.70  & 5.05 \\
Union2.1  & 1.41 & 2.90  & 1.27 \\
Union2.1 + BAO  & 0.18 & 0.16  & 41.25 \\
\hline
\end{tabular}
\caption{Constraints on ($w_0 - w_a$) and ($w_0 - w_{0.5}$). }  
\end{center}
\end{table}

% ################################### Results ################################################################
\section{Results} 

In order to calculate the FoM, we follow Wang (2008) and define ${\rm{FoM}} = {1}/{\sqrt{\det {C(\theta})}}$, 
where $C(\theta)$ is the covariance matrix of the set of parameters $\theta$. Using the prescription of the previous section, we perform 30 realizations of $t(z)$ for each group of $N = 100$, 500, and 1000 data points, with $\sigma_{\rm t}$ = 5\% and 10\%. The central values of the FoM and the corresponding error bars are obtained using a bootstrap method on the original 30 Monte Carlo realizations.

Figures 3 and 4 show the main results of our analysis. The expected FoM is shown as a function of the number of data points for $\sigma_t = 10\%$ (Fig. 3) and $\sigma_t = 5\%$ (Fig. 4). In all panels, solid red lines represent the FoM for the current observational $t(z)$ sample (Simon {\it et al.} 2009) whereas solid blue lines represent the FoM obtained from a magnitude-redshift test using  580 type Ia supernovae (SNe Ia) of the Union2.1 compilation (Suzuki {\it et al}, 2012). We also performed joint analyses involving $t(z)$, SNe Ia, and six measurements of the BAO peak length scale taken from Percival {\it et al.} (2007), Blake {\it et al.} (2011), and Beutler {\it et al.} (2011). Dashed red and blue lines stand for the joint analyses involving $t(z)$ + BAO and SNe Ia + BAO, respectively. The results for these combinations of age, SNe Ia, and BAO data are shown in Table I.

Panels 3a-4a and 3b-4b correspond, respectively, to the results obtained for parameterizations (\ref{eq:w_CPL}) and (\ref{eq:w_c}). Solid squares represent the FoM for each group of age simulated data,  whereas solid circles correspond to the joint analysis with the six BAO data points. There is a slight dependence of the FoM with the increase of $N$ and a clear and substantial gain with the combination with BAO data. For instance, considering age data only and $\sigma_t = 10\%$ we find that increasing the number of data points improves the FoM by a factor of 1.3  - 2.3, whereas the ratio between $\rm{FoM}_{\rm Age + BAO}$ and $\rm{FoM}_{\rm Age}$ increases by a factor of 4 - 9. In Table 2 we summarize the main results of our analysis and put into numbers the results displayed in Figs. 3 and 4.

Another aspect that is worth emphasizing concerns the use of parameterization (\ref{eq:w_c}). Clearly, there is a significant improvement in the FoM for the $w_0 - w_{0.5}$ plane relative to the one for the $w_0 - w_a$. This is an expected result since the former set of parameters is less correlated than the latter one, which is in full agreement with the results discussed by Wang (2008) using SNe Ia + CMB + BAO data. In our analysis we find that the $\rm{FoM}_{\rm Age}$ and $\rm{FoM}_{\rm Age + BAO}$ increase, respectively, by a factor of 4 and 6 relative to the same quantities for parameterization (\ref{eq:w_CPL}). We also observe that a similar result can also be obtained from the current observational data (Table 1). In this case, regardless of whether we consider the age only, or the age + BAO analysis, we obtain an improvement factor for the figure of merit of $\simeq 3$. 

For the sake of completeness, we also performed our analysis assuming $z_c$ in Eq. (\ref{paramW}) to be a free parameter of the model so that the plane $w_{\rm 0} - w_{\rm a}$ becomes completely uncorrelated. The results for this analysis are shown in Figs. 3c and 4c. Compared to the analysis for parameterization (\ref{eq:w_c}), we find an improvement in the $\rm{FoM}_{\rm Age + BAO}$ and $\rm{FoM}_{\rm Age}$ that varies, respectively, by a factor of 1.5 - 3.4 and 2.0 - 4.6 (see Table 3).

\begin{table}[!t]
\begin{center}
\begin{tabular}{llll}
\hline 
Parametrization (\ref{eq:w_CPL})\\  
$\sigma_{\rm t}$ & $N$ & $\rm{FoM}_{\rm Age}$ & $\rm{FoM}_{\rm Age + BAO}$ \\
\hline
10\% & 100  & $1.05 \pm 0.81$ & $9.13 \pm 6.80$ \\ 
& 500 & $1.35 \pm 0.68$ & $10.76 \pm 5.08$ \\    
% \cline{2-6}
& 1000 & $3.13 \pm 2.01$ & $15.84 \pm 7.18$ \\   
% \cline{2-6}
\hline
5\% & 100 & $1.72 \pm 0.95$ & $14.17 \pm 8.05$ \\
% \cline{2-6}  
& 500 & $3.21 \pm 1.99$ & $14.70 \pm 4.36$ \\
% \cline{2-6}
& 1000 & $5.11 \pm  3.01$ & $21.91 \pm 5.23$ \\
\hline
Parametrization (\ref{eq:w_c}) \\
$\sigma_{\rm t}$ & $N$ & $\rm{FoM}_{\rm Age}$ &  $\rm{FoM}_{\rm Age + BAO}$\\
\hline
10\% & 100 & $3.15 \pm 2.44$ & $28.29 \pm 20.93$ \\    
% \cline{2-6}
& 500 & $4.05 \pm 2.05$ & $32.29 \pm 15.26$ \\ 
% \cline{2-6}
& 1000  & $8.55 \pm 5.13$ & $45.90 \pm 21.16$ \\  
% \cline{2-6}
\hline
5\% & 100  & $5.17 \pm 2.87$ & $42.53 \pm 24.14$ \\
% \cline{2-6}
& 500 & $9.64 \pm 5.98$ & $44.11 \pm 13.07$ \\
% \cline{2-6}
& 1000 & $14.73 \pm 8.80$ & $64.13 \pm 16.62$ \\ 
\hline
\end{tabular}
\caption{Figure of merit obtained from each group of simulated dataset for the parametric spaces ($w_0 - w_a$) and ($w_0 - w_{0.5}$).}
\label{tab_simulated_datasets_fom_sigmafom}
\end{center}
\end{table}

\begin{table}[!t]
\begin{center}
\begin{tabular}{llllllll}
\hline 
\hline 
Parametrization (\ref{paramW}) \\
$\sigma_{\rm t}$ & $N$ & $\rm{FoM}_{\rm{Age}}$ &  $\rm{FoM}_{\rm Age + BAO}$ \\
\hline
10\% & 100  &  $4.02  \pm 2.19$  &  $28.57 \pm  13.93$ \\   
% \cline{2-6}
& 500  & $7.73 \pm 2.88$ & $39.44 \pm 14.98$ \\       
% \cline{2-6}
& 1000 & $12.67 \pm 6.50$ & $53.80 \pm 19.43$ \\
% \cline{2-6}
\hline
5\% & 100 & $7.68 \pm 3.53$ & $48.60 \pm 23.77$ \\   
% \cline{2-6}
& 500 & $15.81 \pm 7.10$ &  $51.28 \pm 12.37$ \\
% \cline{2-6}
& 1000 & $22.60 \pm 9.20$ & $74.89 \pm 20.95$ \\  
\hline
\end{tabular}
\caption{The same as in Table (\ref{tab_simulated_datasets_fom_sigmafom}) for the plane ($w_0 - w_{z_{c}}$).}
\label{tab_s}
\end{center}
\end{table}

\section{Conclusions}

Age estimates of high-$z$ objects constitute a complementary probe to distance-based observations such as SNe Ia and BAO measurements. In this paper, we have explored $t(z)$ constraints on the dark energy EoS from two different routes, namely: calculating the relative error in the expansion age as a function of the relative error in the EoS parameters, and performing MC simulations from current observational data.

Using synthetic samples of $t(z)$ with different sizes and accuracy and their combinations with BAO data, we have found a significant improvement in the FoM for the planes $w_0 - w_{0.5}$ and $w_0 - w_{z_{c}}$ relative to the one for the $w_0 - w_a$ space. We have also studied the dependence of the figure of merit with the number of data points $N$, with $\sigma_t$, as well as with the combination of $t(z)$ and BAO observations,  and found that the latter two provide the more substantial gains. One of the results of our analysis is that $t(z)$ data may become competitive with SNe Ia observations only for $\sigma_t < 5\%$. This result certainly reinforces the importance of a better understanding of the systematic errors in the age determination of high-$z$ objects as important probes to the late stages of the Universe.

%###########################Ackowledgements#########################

\acknowledgements

The authors thank CAPES, CNPq, and FAPERJ for the grants under which this work was carried out.

%################### References ###################################

\label{lastpage}
\end{document}